\documentclass[prl,twocolumn,superscriptaddress,citeautoscript,amsart,longbibliography,nofootinbib]{revtex4-2}

\usepackage{graphicx}
\usepackage{multirow}
\usepackage{color}
\usepackage{bm}
\usepackage{times}
\usepackage{amsmath,bm,amsfonts}
\usepackage{dcolumn}
\usepackage{graphicx}
\usepackage{latexsym}
\usepackage{hhline}
\usepackage{braket}
\usepackage{bbold}
\usepackage{multirow}

\def\ket#1{|#1\rangle }

\def\d{\partial}

\begin{document}
\title{
Superconductivity-induced spectral weight transfer due to quantum geometry
}
\author{Junyeong \surname{Ahn}}
\email{junyeongahn@fas.harvard.edu}
\affiliation{Department of Physics, Harvard University, Cambridge, MA 02138, USA}

\author{Naoto \surname{Nagaosa}}
\email{nagaosa@riken.jp}
\affiliation{RIKEN Center for Emergent Matter Science (CEMS), Wako, Saitama 351-0198, Japan}
\affiliation{Department of Applied Physics, The University of Tokyo, Bunkyo, Tokyo 113-8656, Japan}

\begin{abstract}
Optical spectral weight transfer associated with the onset of superconductivity at high energy scales compared with the superconducting gap has been observed in several systems such as high-$T_c$ cuprates.
While there are still debates on the origin of this phenomenon, a consensus is that it is due to strong correlation effects beyond the BCS theory.
Here we show that there is another route to a nonzero spectral weight transfer based on the quantum geometry of the conduction band in multiband systems.
We discuss applying this idea to the cuprates and twisted multilayer graphene.
\end{abstract}

\date{\today}

\maketitle

{\it Introduction.---}
Superconductivity is a phenomenon due to the binding electron pairs near the Fermi level.
A natural expectation is that any physical phenomenon at energy scales much larger than the pairing gap is unaffected by superconductivity.
However, experiments since the 1990s have shown that cuprate superconductors are exceptions.
Cuprate superconductors showed a change in the optical spectrum even at an energy scale much larger (about 100 times) than the pairing gap size during the superconducting transition~\cite{fugol1993temperature,fugol1993anomalies,basov1999sum,rubhausen2001superconductivity,molegraaf2002superconductivity,deutscher2005kinetic,gedik2005abrupt,carbone2006doping,giannetti2011revealing}.
Since the optical spectral sum over all energies does not change by the $f$-sum rule, the decreased (increased) spectral weight at high energies is transferred to (from) low energies.
Underdoped cuprates showed spectral weight transfers from high to low energies~\cite{fugol1993temperature,fugol1993anomalies,basov1999sum,rubhausen2001superconductivity,molegraaf2002superconductivity,boris2004plane,deutscher2005kinetic,gedik2005abrupt,carbone2006doping,giannetti2011revealing}, while overdoped cuprates showed the opposite~\cite{molegraaf2002superconductivity,deutscher2005kinetic,gedik2005abrupt,carbone2006doping,giannetti2011revealing}, which was reproduced in dynamical mean-field calculations~\cite{gull2012energetics,fratino2016organizing}.
The origin of the superconductivity-induced spectral weight transfer (also called color change~\cite{hirsch1992superconductors,hirsch2002true} or UV-IR mixing~\cite{phillips2010colloquium}) has been debated over 20 years and continues up to date~\cite{hirsch1992apparent,hirsch1992superconductors,hirsch2002true,marsiglio2006intraband,phillips2010colloquium,phillips2020exact}.

The anomalous spectral weight transfer is thought to be closely related to the exotic pairing mechanism in the cuprates, beyond the Bardeen-Cooper-Schrieffer (BCS) theory.
For example, an issue discussed intensively is whether the high-temperature superconductivity in the cuprates is driven by the interaction energy or by the kinetic energy~\cite{basov2005electrodynamics,lee2006doping}.
In BCS theory, superconductivity decreases the interaction energy while increasing the kinetic energy.
As the low-energy spectral weight is approximately given by the minus of the kinetic energy, the superconductivity-induced spectral weight transfer from low to high energies reported in overdoped cuprates~\cite{deutscher2005kinetic,gedik2005abrupt,carbone2006doping,giannetti2011revealing} is thought to be explained within BCS theory.
On the other hand, the spectral weight transfer from high to low energies implies that the kinetic energy of the conduction electrons is reduced by superconductivity~\cite{hirsch1992apparent,hirsch1992superconductors}.
This led to a proposal that the cuprates are kinetic-energy-driven superconductors in the underdoped and overdoped regime~\cite{hirsch1992superconductors}.

Another related issue is the role of Mottness~\cite{lee2006doping,phillips2010colloquium}.
Recently, Phillips, Yao, and Huang (PYH)~\cite{phillips2020exact} studied superconductivity in doped Mott insulators by adding superconducting pairing to an exactly solvable Model due to Hatsugai and Kohmoto~\cite{hatsugai1992exactly}.
Interestingly, they obtained a superconducting state distinguished from the conventional BCS state.
Among the properties of the Mottness-induced non-BCS superconductivity, one particularly intriguing property is the superconductivity-induced spectral weight transfer.
They put the chemical potential on the lower Hubbard band and calculated the integrated single-particle spectral function for the lower Hubbard band.
Remarkably, the spectral sum increased when superconductivity sets in, which, by a $f$-sum rule stating that the total spectrum sum of the lower and upper Hubbard band is invariant, indicates that the spectral weight was transferred from the upper to the lower Hubbard band. 
Based on this observation, PYH concluded that Mottness is the key to the problem of anomalous spectral weight transfer in cuprate superconductors.

In this Letter, we point out that a superconductivity-induced optical spectral weight transfer from high to low energies appears naturally within the multiband BCS theory.
What is central to the problem is instead the so-called quantum geometric effect, which has been considered mostly in the context of flat-band superconductivity~\cite{peotta2015superfluidity,hu2019geometric,julku2020superfluid,xie2020topology}.
Although this effect in unconventional superconductors is not so significant as in flat-band systems, it can be large enough to have an imprint on experimental observations, as in the cuprates.
Our result shows the importance of quantum geometry in understanding the exotic properties of unconventional superconductors.

\begin{figure*}[t!]
\includegraphics[width=0.8\textwidth]{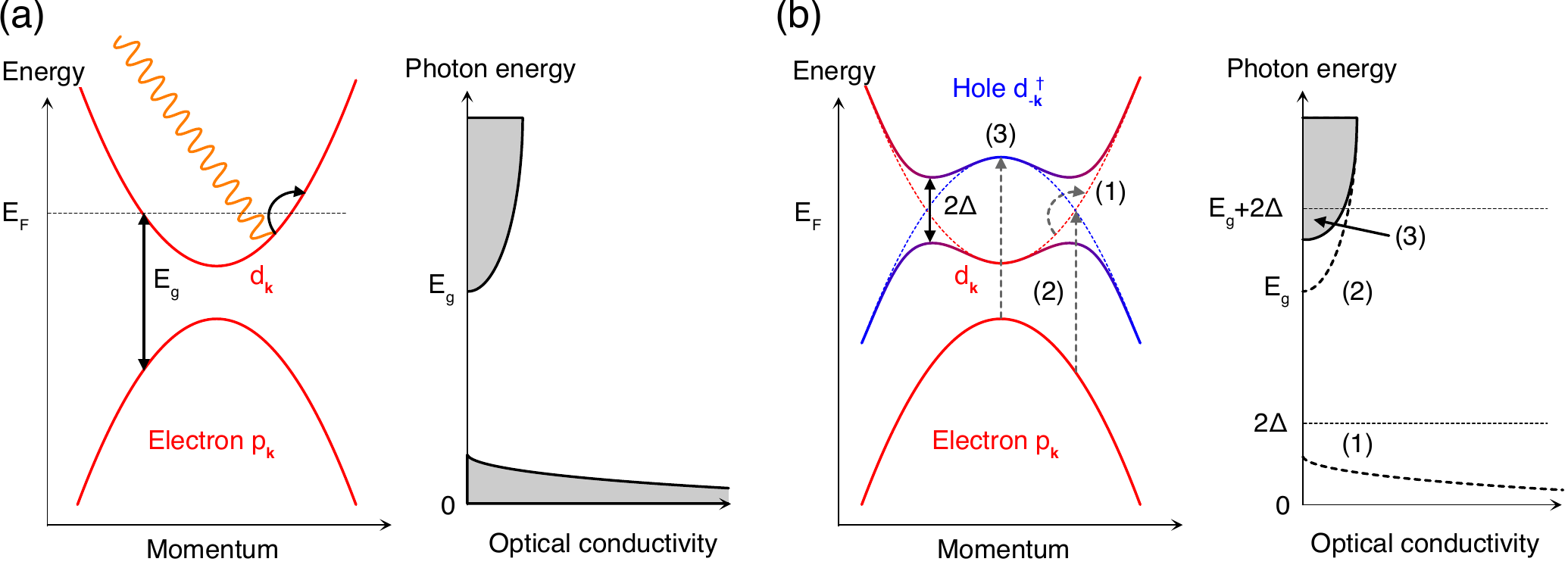}
\caption{
A simplified picture of the optical spectrum of metals in the normal and superconducting states.
(a), Normal state.
$E_g$ represents the energy of the excitation to the Fermi surface.
There may be additional excitation channels to the Fermi surface from lower energy levels, or from the Fermi surface to higher energy levels, both which we do not show here for simplicity.
$p$ and $d$ represent two different electronic bands. 
(b), Superconducting state.
$2\Delta$ indicates the superconducting gap.
Both (1) intraband and (2) interband spectral weights are lost by the opening of the superconducting gap, while (3) new electron-hole mixing excitation channels opens up.
}
\label{fig:spectrum}
\end{figure*}

{\it Optical f-sum rules.---}
Let us begin by explaining the optical spectral weight transfer.
The optical frequency sum rule states that the oscillator strength, the spectrum sum of the optical conductivity tensor over all frequencies, depends only on the density of electrons $n$.
\begin{align}
\label{eq:f-sum}
\frac{2}{\pi}\int^{\infty}_0 d\omega \sigma^{ca}_{1}(\omega)
=\frac{ne^2}{m_e}\delta^{ca},
\end{align}
where $\sigma_1$ is the real part of the conductivity tensor, $-e$ is the electron charge, $m_e$ is the bare electron mass, and $\delta^{ca}$ is the Kronecker delta.
This $f$-sum rule is independent of the form of the interactions and the form of the ground state.
Combining this sum rule with the conservation of charge, one obtains the Ferrel-Glover-Tinkham (FGT) sum rule for the superfluid weight ${\cal D}^{ca}\equiv(2/\pi)\int^{0^+}_{0} d\omega\sigma^{ca}_{1s}(\omega)$,
which reads
\begin{align}
\label{eq:FGT}
{\cal D}^{ca}
&=
\frac{2}{\pi}\int^{\infty}_{0^+} d\omega\left[\sigma^{ca}_{1n}(\omega)-\sigma^{ca}_{1s}(\omega)\right],
\end{align}
where we use $\int^{0^+}_{0} d\omega\sigma^{ca}_{1n}(\omega)=0$, and the subscripts $n$ and $s$ indicate the normal and superconducting states.
The spectral sum of the optical conductivity tensor is thus an important factor determining the superfluid response, and this is what we call the optical spectral weight.
The missing spectral weight of the ``regular" optical conductivity (where regular meaning $\omega>0$) appears in the superfluid weight.

{\it Optical spectral weight transfer.---}
In conventional superconductors, the missing spectral weight appears mostly near the Fermi level (photon energy up to the order of the pairing gap) because the Drude weight is lost by the opening of the superconducting gap.
This is natural given that superconducting pairing occurs near the Fermi level.

However, we note that, after the electrons at the Fermi level condense to become Cooper pairs, it requires additional energy to excite those electrons {\it for any excitation channel} to overcome the binding energy --- this is one of the main points of this Letter.
Therefore, a missing spectral weight always appears for interband transitions also, which may occur at very high energies [Fig.~\ref{fig:spectrum}(b)].
Thanks to this spectral weight transfer, even exactly flat bands with zero Drude weight can have nonzero superfluid weight~\cite{peotta2015superfluidity}.
This interband superfluid weight was recently proposed to be relevant in the superconductivity of flat bands in twisted bilayer graphene~\cite{hu2019geometric,julku2020superfluid,xie2020topology}, although it was not discussed in the context of the spectral weight transfer.

There are also electron-hole mixing optical excitation channels generated by superconducting pairing [Fig.~\ref{fig:spectrum}(b)].
These excitations occur mainly between electrons and holes originating from different normal-state bands, because particle-hole-symmetric excitations are usually forbidden by inversion symmetry in superconductors~\cite{ahn2021theory,ahn2021many}.
The emergence of these new excitations transfers the optical spectral weight from low to high energies, reducing the superfluid weight.

We use a model of Dirac fermion
\begin{align}
\label{eq:Dirac}
h=\hbar v_F(k_x\sigma_x+k_y\sigma_y)
\end{align}
to show the spectral weight transfer explicitly.
Let us note that an effective model does not satisfy the sum rule Eq.~\eqref{eq:f-sum} in general.
Instead, the spectrum sum is $(2/\pi)\int^{\infty}_0 d\omega \sigma^{ca}_{1}(\omega)
=(e^2/\hbar^2)\int_{\bf k}\d_{k^c}\d_{k^a}h,$ which may change by superconductivity even when the electron density is conserved.
The physical meaning behind this is that the spectral weight may be transferred from the bands not included in the model.
However, our linearized Dirac model preserves the spectrum sum because $\d_{k^c}\d_{k^a}h=0$, such that the spectral sum depends only on the regularization at the high-energy cut-off $\Lambda$, which is not affected by superconductivity in the limit of $\Lambda\rightarrow \infty$.
Therefore, an optical spectral weight transfer occurs only between the two bands in the model.

\begin{figure}[t!]
\includegraphics[width=8.5cm]{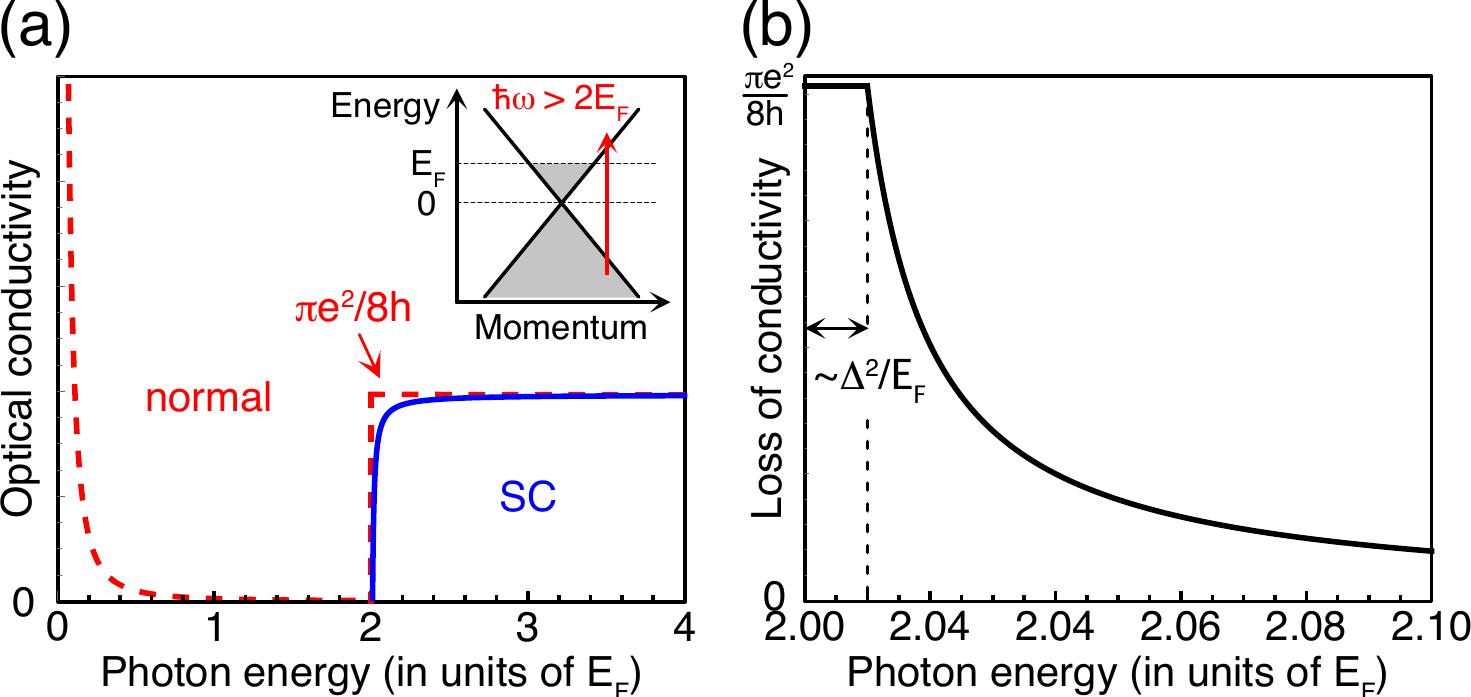}
\caption{
Optical conductivity of a Dirac fermion in normal and superconducting states in two dimensions.
(a) Optical conductivity.
Only regular (non-superfluid) parts are shown.
The red dashed line and solid blue line correspond to the normal and superconducting states, respectively.
The Drude conductivity is shown with a finite inverse scattering rate of $\Gamma=0.01E_F$ for its visual presentation, while we take the clean limit $\Gamma\rightarrow 0$ to calculate the interband conductivities.
We consider an $s$-wave gap with $\Delta=0.1E_F$ in the superconducting state.
The inset shows the spectrum of the Dirac fermion in the normal state.
(b) Loss of the interband optical conductivity by the superconducting transition.
All conductivities are calculated at zero temperature.
}
\label{fig:Dirac-spectra}
\end{figure}

The optical conductivity in the normal state is
\begin{align}
\sigma^{ca}_{1n}(\omega)
&=\delta_{ca}\frac{e^2}{h}\left(\frac{\Gamma}{\omega^2+\Gamma^2}\frac{E_F}{2\hbar}+\frac{\pi}{8}\Theta(\omega-2E_F)\right),
\end{align}
where we introduce a finite relaxation rate $\Gamma$ for the intraband (Drude) response, while neglecting the relaxation for interband transitions, and $\Theta$ is the Heaviside step function.
In the superconducting state with an $s$-wave gap $2\Delta$, it reduces to
\begin{align}
\sigma^{ca}_{1s}(\omega)
&=\frac{\pi}{2}{\cal D}^{ca}\delta(\omega)+\delta_{ca}\frac{e^2}{h}\frac{\pi}{8}F(\omega)\Theta(\omega-2\sqrt{\mu^2+\Delta^2}),
\end{align}
where
\begin{align}
{\cal D}^{ca}
=\delta^{ca}\frac{e^2}{h}\frac{E_F}{h}\left(\frac{E_F}{\mu}\right),
\end{align}
and the Fermi level $E_F$ in the normal state is renormalized to $\mu$ in the superconducting state to preserve the average electron density, which satisfies the self-consistency equation
\begin{align}
E_F^2=\mu\sqrt{\mu^2+|\Delta|^2}+|\Delta|^2\tan^{-1}\frac{\mu}{\sqrt{\mu^2+|\Delta|^2}},
\end{align}
$F(\omega)=(k_{\omega}/\omega \mu)(\sqrt{\Delta^2+(k_{\omega}+\mu)^2}-\sqrt{\Delta^2+(k_\omega-\mu)^2})$, and $k_{\omega}=\omega\sqrt{\omega^2-4(\mu^2+\Delta^2)^2}/\sqrt{\omega^2-(2\mu)^2}$ is the magnitude the momentum where optical transition occurs.
Figure~\ref{fig:Dirac-spectra} shows the regular part of the optical conductivity in the normal and superconducting states.
Both intraband and interband optical conductivities originating from the electrons near the Fermi level get reduced when those electrons form Cooper pairs [Fig.~\ref{fig:Dirac-spectra}(a)].

It is worth noting a difference between the intraband and interband contributions.
While all the intraband optical spectral weight is lost over the gap size $0<\hbar\omega<2\Delta$, the interband weight is lost only partially over the gap size $E_g<\hbar\omega<E_g+2\Delta$, where $E_g=2E_F$ is the optical interband spectral gap in the normal state, and $E_F$ is the Fermi energy measured from the Dirac node [Fig.~\ref{fig:Dirac-spectra}(b)].
The is because new optical transition channels are created at $\hbar\omega\sim E_g=2E_F$ in the superconducting state by the electron-hole mixing, compensating the loss of the spectral weight in the transition channels of the normal state.
In contrast, electron-hole-mixing transitions are usually prohibited at the superconducting gap scale~\cite{mahan2013many,xu2019nonlinear,ahn2021theory,ahn2021many}, such that most of the intraband weight for $\hbar\omega<2\Delta$ transforms to the superfluid weight when $\Gamma\ll 2\Delta$.

The interband contributions arise as corrections of order $(\Delta/E_F)^2$ to the intraband superfluid weight in the clean limit.
This is because the interband optical conductivity is $O(E_F^{d-2})$ such that interband superfluid weight is $O(\Delta^2E_F^{d-3})$, while the intraband contribution is the Drude weight --- scaling as $O(E_F^{d-1})$.
In the case of the Dirac model in Eq.~\eqref{eq:Dirac}, an additional logarithmic correction appears: the ratio between interband and intraband contributions is $(\Delta/E_F)^2\log(E_F/\Delta)$ for small $\Delta/E_F$.
As conventional superconductors have a very small ratio $\Delta/E_F\lesssim10^{-4}$, the interband contribution is negligible for them.
However, unconventional superconductors can reach much higher value of $\Delta/E_F\gtrsim 0.1$~\cite{shibauchi2020exotic,park2021tunable,hao2021electric,nakagawa2021gate}, such that the interband contribution is a few percent.

\begin{figure*}[t!]
\includegraphics[width=\textwidth]{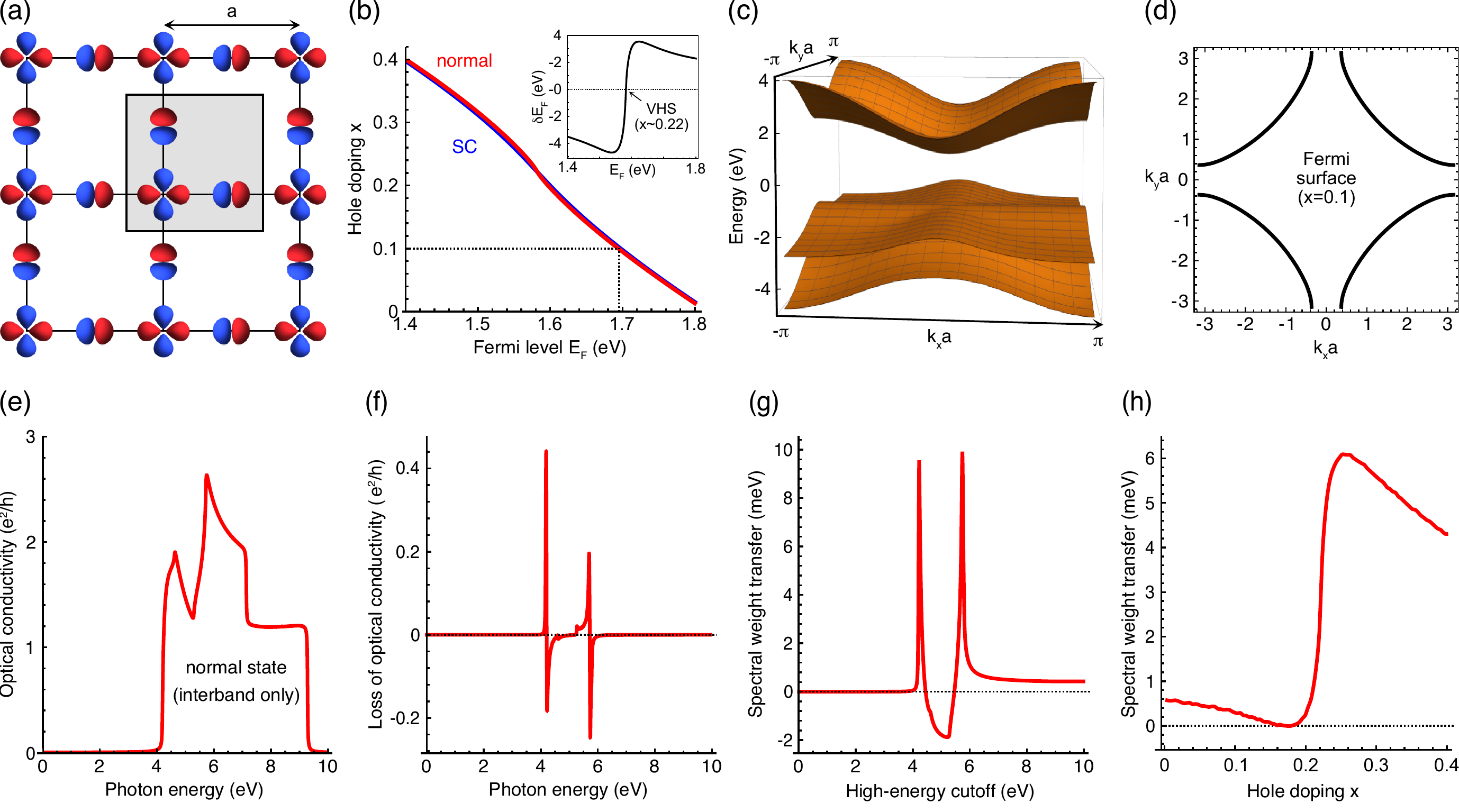}
\caption{
Superconductivity-induced spectral weight transfer in a model of the CuO$_2$ plane in cuprate superconductors.
(a-d) Tight-binding model.
(a) Location of $p_{x,y}$ and $d_{x^2-y^2}$ orbitals on the Lieb lattice.
The unit cell is shown as a shaded rectangle.
We consider the tight-binding model on this lattice with hopping up to the next-nearest neighbor Eq.~\eqref{eq:Lieb}.
We take onsite energies $\epsilon_d=0$ eV and $\epsilon_p=-2.416$ eV and hopping amplitudes $t_{pp}=751$ meV and $t_{pd}=1.257$ eV.
(b) Hole doping $x=1-n$ as a function of Fermi level $E_F$.
$n$ is the number of conduction electron per unit cell.
The red and blue curve corresponds to the normal and superconducting states, where $\Delta_d=20$ meV in the latter.
We take $E_F=1.695$ eV in (d-g).
This choice corresponds to hole doping $x=0.1$ per unit cell from the half-filled conduction band.
The inset shows $\delta E_F=E_F^{\rm SC}-E_F^{\rm normal}$ needed to preserve the average charge, which changes sign at the van Hove singularity (VHS).
(c) Band structure.
(d) Fermi surface of a hole pocket centered at $(\pi,\pi)$.
(e) Interband optical conductivity in the normal state.
Only the longitudinal part $\sigma^{xx}=\sigma^{yy}$ is shown.
The contribution from the intraband Drude conductivity is not included here.
(f) Superconductivity-induced reduction of the interband optical conductivity.
We take $\Delta_d=20$ meV.
(g) Spectral weight transfer corresponding to (e).
For numerical calculations of the optical conductivity, we take finite broadening $\hbar\Gamma=10$ meV for the Lorentzian.
}
\label{fig:model}
\end{figure*}

In disordered systems, the interband contributions are relatively enhanced~\cite{hirsch2000where}.
Disorder significantly reduces the intraband contribution to the superfluid weight when the disorder-induced broadening $\hbar\Gamma$ of the Drude spectrum is comparable or larger than the superconducting gap $2\Delta$, because the spectral weight below $\hbar\omega<2\Delta$ is reduced [Fig.~\ref{fig:spectrum}(b)].
However, the interband contribution is not affected much by disorder as long as the interband optical spectral gap is larger than the spectral broadening.
Therefore, the interband contribution is relatively much enhanced for $2\Delta\lesssim\hbar\Gamma<E_g$.

{\it Three-orbital model of the CuO$_2$ plane.---}
Let us apply the above idea to a model of cuprate superconductors.
We consider the three-orbital tight-binding model on the Lieb lattice [Fig.~\ref{fig:model}(a)] described by
\begin{align}
\label{eq:Lieb}
h=
1_{\rm spin}\otimes
\begin{pmatrix}
\epsilon_d & f(k_x)&-f(k_y)\\
 f^*(k_x)& \epsilon_p &g(k_x,k_y)\\
-f^*(k_y)&g^*(k_x,k_y)& \epsilon_p\\
\end{pmatrix}
\end{align}
in the basis of $(\uparrow,\downarrow)$ spins $\otimes$ $(d_{x^2-y^2},p_x,p_y)$ orbitals, $f(k)=t_{pd}(1-e^{ika})$, $g(k_x,k_y)=-t_{pp}(1-e^{-ik_xa})(1-e^{ik_ya})$, and $a$ is the lattice constant.
We take $\epsilon_d=0$ eV, $\epsilon_p=-2.416$ eV, $t_{pp}=751$ meV, and $t_{pd}=1.257$ eV, which are proper values to model cuprate superconductor HgBa$_2$CuO$_4$~\cite{hirayama2018ab}.
We consider the effect of doping by tuning only the Fermi level $E_F$ with other parameters fixed.
We first take $E_F=1.695$ eV, corresponding to the hole doping $x=1-n\sim 0.1$ in the underdoped regime, where $n$ is the number of conduction electrons per unit cell [Fig.~\ref{fig:model}(b)].

Figure~\ref{fig:model}(c,d,e) shows the band structure, Fermi surface, and interband optical conductivity in the normal state.
We neglect the Drude conductivity and focus on the interband contributions.
As we do not include correlation effects, the band structure and optical conductivity differ significantly from the experimental observations in high-$T_c$ cuprates.
Here, we only aim to show that spectral weight can be transferred from high to low energies without correlation effects.

We consider an orbital-independent pairing function $\Delta_{\bf k}=\Delta_d(\cos k_y-\cos k_x)$, such that the gap of $2\Delta_{\bf k}$ opens at the Fermi level.
Figure~\ref{fig:model}(f) shows the reduction of the interband optical conductivity for $\Delta_d=20$ meV.
Most of the change occurs at around $4$ eV and $6$ eV, which are 100 and 150 times larger than the gap $2\Delta_d$, respectively.
This change is due to the modification of the excitations to the Fermi level from the two occupied $p$ orbitals.
The corresponding spectral weight transfer up to high-energy cutoff $\hbar\omega_c$, ${\rm SWT}(\omega_c)=(h/e^2)(2/\pi)\int^{\omega_c}_{0} d\omega (\sigma^{xx}_{1n}-\sigma^{xx}_{1s})$, is shown in Fig.~\ref{fig:model}(g) in unit of meV.
Here, the total spectral weight transfer $0.42$ meV is comparable to $1$ meV observed in the underdoped regime~\cite{molegraaf2002superconductivity,deutscher2005kinetic,gedik2005abrupt,carbone2006doping,giannetti2011revealing}.

The spectral weight transfer from high to low energies occurs for most other doping levels also [Fig.~\ref{fig:model}(h)].
Moreover, our data reproduce the decrease of the spectral weight transfer observed in the cuprates as $x$ increases from the underdoped regime~\cite{deutscher2005kinetic,gedik2005abrupt,carbone2006doping,giannetti2011revealing}, although the sign change observed in the overdoped regime is almost invisible in our model calculation.

{\it Quantum geometric perspective.---}
The optical spectral weight transfer --- equivalently, by the FGT sum rule, the interband contribution to the superfluid weight --- has an interesting quantum geometric interpretation.
The interband contribution to the superfluid weight is often called a ``quantum geometric" contribution because it appears as the Fubini-Study metric of the conduction band in the exactly flat-band limit~\cite{peotta2015superfluidity}.
Namely, $D^{ca}\propto {\rm Re}\int_{\bf k}Q_{ca}$, where $Q_{ca}=\sum_{n\in {\rm flat},m\in {\rm remote}}\braket{n|i\d_{k^c}|m}\braket{m|i\d_{k^a}|n}$ is the Fubini-Study metric of the flat bands.
Recent studies by the present authors~\cite{ahn2020low,ahn2021riemannian} allow a stronger quantum geometric interpretation that the interband superfluid weight is quantum geometric even away from the flat-band limit.
We showed that the matrix elements in the interband optical conductivity tensor have the form of a quantum metric in the normal state;
\begin{align}
\sigma^{ca}(\omega)
=\frac{\omega e^2}{2}\sum_{m,n}\int_{\bf k} \delta\left(E_m-E_n-\hbar\omega\right)f_{nm}Q^{mn}_{ca},
\end{align}
where $\int_{\bf k}=\int d^dk/(2\pi)^d$ in $d$ spatial dimensions, $f_{nm}=f_n-f_m$, and $f_n$ is the Fermi-Dirac distribution of the electronic single-particle state $\ket{n}$.
Here, $Q^{mn}_{ca}=\braket{n|i\d_{k^c}|m}\braket{m|i\d_{k^a}|n}$ is the quantum metric for the pair of states $\ket{m}$ and $\ket{n}$~\cite{ahn2021riemannian}, which produces the Fubini-Study metric after summed over $m$ and $n$.
The FGT sum rule then implies that the interband superfluid weight is related to the quantum metric in the normal state.
For a pairing that is uniform for all orbital degrees of freedom, one can see this explicitly in the expression derived in Ref.~\cite{liang2017band}, which we rewrite in terms of the quantum metric as
\begin{align}
\label{eq:SF-inter}
{\cal D}^{ca}_{\rm inter}
=\frac{e^2}{\hbar^2}\sum_{m\ne n}\int_{\bf k}\left(\frac{1}{E_m}-\frac{1}{E_n}\right)\frac{\Delta^2(\xi_n-\xi_m)}{\xi_n+\xi_m}{\rm Re}[Q^{mn}_{ca}],
\end{align}
where $\xi_n=\epsilon_n-\mu$, $\epsilon_n$ is the energy eigenvalue of the state $\ket{n}$ in the normal state, and $E_n=\sqrt{\xi_n^2+\Delta^2}$.
An orbital-dependent pairing can complicate the quantitative relation between quantum geometry and superfluid weight~\cite{liang2017band} because then the optical conductivity in the superconducting state does not simply depend on the quantum metric in the normal state.
However, even in this case, the geometric perspective can still be useful for a qualitative understanding.

Quantum geometry gives fresh insight into the single-band effective models of the superconductivity-induced spectral weight transfer.
In Ref.~\cite{hirsch1992apparent}, Hirsch noted that, if we consider the non-orthogonality of the atomic orbitals, repulsive interactions need to be included on the bond sites as well as the atomic sites in the Hubbard model.
When superconductivity occurs, the bond-site interaction increases the total spectral weight, the sum of the regular and superfluid weight~\cite{hirsch1992apparent}.
Such a non-conservation of the spectral sum is allowed in effective models and is interpreted as the spectral weight transfer from remote bands, not included in the model.
We can understand the importance of the bond-site interaction by noting that the spreading of the Wannier function is characterized by the Fubini-Study metric~\cite{marzari2012maximally,peotta2015superfluidity}.
As the optical spectral weight transfer is characterized by a quantum metric, intimately related to the Fubini-Study metric, it is also related to the spreading of the Wannier function.
In Hirsch's model, atomic orbitals take the role of the Wannier function, so their overlap due to the spreading is responsible for the spectral weight transfer.

{\it Discussion.---}
Our general arguments show that the superconductivity-induced optical spectral weight transfer can be understood as quantum geometric effects within the multiband BCS theory.
This is essentially the same phenomenon as the geometric superfluidity studied previously~\cite{peotta2015superfluidity,liang2017band}.
However, our approach has practical advantages over the previous approach as well as conceptual merits in the understanding of the physics of the cuprates.

Our approach provides a promising way to measure the geometric contribution to the superfluid weight in flat-band systems through the superconductivity-induced optical spectral weight transfer.
It was theoretically proposed that the geometric contribution can take a significant portion in flat-band systems~\cite{peotta2015superfluidity,liang2017band} such as twisted bilayer~\cite{hu2019geometric,julku2020superfluid,xie2020topology} and trilayer~\cite{park2021tunable,hao2021electric} graphene.
However, this proposal has not been tested experimentally.
Since the optical spectral gap between the flat bands and the other bands is an order of $10\;{\rm meV}$ at the first magic angle (twisting angle of $\sim 1.1^{\circ}$) in twisted bilayer graphene, the optical spectrum needs to be measured down to the terahertz scale.
This measurement below superconducting transition temperature $T_c\sim 1K$ is currently challenging but may be possible in the near future.
For the trilayer, one would need to measure the spectrum down to lower energies, because there will be additional geometric contributions from the Dirac cones coexisting with the flat bands.

In principle, optically measuring the geometric superfluid weight requires the knowledge of the optical conductivity at arbitrarily high photon energies.
However, in practice, one can achieve high precision with a moderate high-energy cutoff.
This is because the quantum metric $Q^{mn}_{ba}$ is be suppressed for a large energy difference between $m$ and $n$ bands.
To see this, note from Eq.~\eqref{eq:SF-inter} that ${\cal D}_{\rm inter}\sim (\Delta^2/E_F)\sum_{m\ne n}\int_{\bf k}{\rm Re}Q^{mn}$ because the transitions from or to the Fermi level are dominant contributions.
Let us suppose that $n$ is the unique metallic band, and then the sum $\sum_{m\ne n}$ runs over $m$ only.
We have $Q^{mn}_{ba}=v^b_{nm}v^a_{mn}/\omega_{mn}^2\sim O(\omega_{mn}^{-2})$ for large energy difference $\hbar\omega_{mn}$, because the interband velocity matrix element $v^a_{mn}=\braket{m|\hat{v}^a|n}$ does not diverge for $\omega_{mn}\rightarrow \infty$.
If we consider the density of states $\int_{\bf k}\sum_{m:\omega_{mn}\sim\omega}\propto \omega^{d/2-1}$ in $d$ spatial dimensions at $\omega_{mn}=\omega$, which is the case for transitions to free electron states, the contribution at $\omega$ is suppressed by a factor $1/\omega^{(6-d)/2}$ at large $\omega$.

Although we emphasize the role of quantum geometry, we should also note that disorder and correlation effects need to be seriously considered to explain all the salient features appearing in the cuprates.
For instance, in the single-particle-band picture, the spectral weight transfer appears within a small energy window comparable to the pairing gap for each transition channel.
On the other hand, the spectral weight transfer in the cuprates occurs over a very broad range of over an electron-volt scale.
The incoherent nature of the electronic states is crucial for broadening.
Also, there is a finite optical spectral weight between the upper and lower Hubbard bands in the cuprates.
This can lead to a superconductivity-induced optical spectral weight transfer at the Mott gap scale, larger than the charge transfer gap.
Finally, we note that a concrete quantum geometric interpretation of the optical conductivity and superfluid weight is yet to be developed for general strongly correlated systems.

{\it Note added.---}
While finalizing this manuscript, we have noticed a closely related recent work by Chen and Huang~\cite{chen2021probing}.
They also study superconductivity-induced change in the optical spectrum, although its relation to the superfluid weight is not discussed.
After we upload our preprint on arXiv, we became aware of related works by Hazra, Verma, and Randeria~\cite{hazra2019bounds,verma2021optical}, where geometric superfluidity is related to optical conductivity for flat-band systems.

\begin{acknowledgments}
J.A. thanks Ashvin Vishwanath for encouraging him to publish this work.
We appreciate Ashvin Vishwanath, Philip Kim, and Matteo Mitrano for helpful comments.
J.A. was supported by the Basic Science Research Program through the National Research Foundation of Korea (NRF) funded by the Ministry of Education (Grant No. 2020R1A6A3A03037129) and by the Center for Advancement of Topological Semimetals, an Energy Frontier Research Center funded by the US Department of Energy Office of Science, Office of Basic Energy Sciences, through the Ames Laboratory under contract No. DE-AC02-07CH11358.
N.N. was supported by JST CREST Grant Number
JPMJCR1874 and JPMJCR16F1, Japan, and JSPS KAKENHI
Grant number 18H03676.
\end{acknowledgments}


%

\end{document}